\documentclass[a4paper]{article}

\usepackage[latin1]{inputenc} %
\usepackage[T1]{fontenc} %
\usepackage{RR}
\usepackage{hyperref}

\usepackage{graphicx,amsmath}
\usepackage{subfigure}
\usepackage{epsfig}
\usepackage{verbatim}
\usepackage{algorithm}
\usepackage{algorithmic}
\usepackage{color}
\usepackage{multirow}
\usepackage{latexsym}

\newtheorem{thm}{Theorem}

\newtheorem{rem}{Remark}

\newcommand{\beq}{\begin{equation}}
\newcommand{\eeq}{\end{equation}}

\def\t1{\hbox{\bf 1}}
\def\P{\hbox{\sf P}}
\def\E{\hbox{\sf E}}

\RRdate{July 2010}

\RRauthor{%
Konstantin Avrachenkov\thanks{INRIA Sophia Antipolis-M\'editerran\'ee, France, k.avrachenkov@sophia.inria.fr}
\and
Evsey Morozov\thanks{Institute of Applied Mathematical Research, Karelian Research Centre RAS and Petrozavodsk University,
Russia, emorozov@karelia.ru}
}

\authorhead{K. Avrachenkov \& E. Morozov}

\RRtitle{L'analyse de la stabilité de la file d'attente de type GI/G/c/K
avec des clients qui reviennent à un taux constant}
\RRetitle{Stability Analysis of GI/G/c/K Retrial Queue with Constant Retrial Rate}
\titlehead{Stability Analysis of GI/G/c/K Retrial Queue with Constant Retrial Rate}

\RRresume{
On considère une file d'attente de type GI/G/c/K avec des clients qui reviennent à un taux constant.
Le système se compose d'une file d'attente primaire et une file d'attente orbite. La file d'attente
primaire a $c$ serveurs identiques et peut accueillir le nombre maximal de $K$ clients.
Si un arrivé trouve la file d'attente primaire pleine, il rejoint l'orbite.
Les clients qui entrent dans le système pour la première fois arrivent selon un processus de renouvellement.
Les clients ont un temps de service générale iid. Les clients dans la file d'attente orbite
essaient d'entrer dans la file d'attente primaire après un temps avec une distribution exponentielle
indépendante de la longueur de la file d'attente orbite. Les commutateurs téléphoniques,
le contrôle d'accès au support, et les courte transferts TCP sont quelques-unes des applications de
le système étudié. Pour ce système, nous établissons les conditions de stabilité suffisantes.
Notre modèle est très général. En plus des cas particuliers (par exemple, M/G/1/1 ou M/M/c/c),
le modèle proposé couvre les cas particuliers du modèle de service déterministe et le modèle
Erlang avec des clients qui reviennent. Les derniers cas particuliers n'ont pas été
considéré dans le passé. Les conditions de stabilité obtenus ont une interprétation probabiliste
tres claire.
}

\RRabstract{
We consider a GI/G/c/K-type retrial queueing system with constant retrial
rate. The system consists of a primary queue and an orbit queue. The primary
queue has $c$ identical servers and can accommodate the maximal number of $K$ jobs.
If a newly arriving job finds the full primary queue, it joins the orbit.
The original primary jobs arrive to the system according to a renewal process.
The jobs have general i.i.d. service times. A job in front of the orbit queue
retries to enter the primary queue after an exponentially distributed time
independent of the orbit queue length. Telephone exchange systems, Medium
Access Protocols and short TCP transfers are just some applications of the
proposed queueing system. For this system we establish minimal sufficient
stability conditions. Our model is very general.  In addition, to the known
particular cases (e.g., M/G/1/1 or M/M/c/c systems), the proposed model
covers as particular cases the deterministic service model and the Erlang
model with constant retrial rate. The latter particular cases have not been
considered in the past. The obtained stability conditions have clear
probabilistic interpretation.
}

\RRmotcle{File d'Attente avec des Clients qui Reviennent, Stabilité Stochastique,
Processus de Renouvellement, Regeneration}
\RRkeyword{Retrial Queue, Constant Retrial Rate, Renewal Process, Regeneration,
Stochastic Stability}

\RRprojet{Maestro}  
\RRtheme{\THCom} 
\URSophia 

\begin{document}

\RRNo{7335}

\makeRR

\section{Introduction}

We consider a GI/G/c/K-type retrial queueing system. The system
consists of a primary queue and an orbit queue. The primary queue
has $c$ identical servers and can accommodate the maximal number of
$K$ jobs. If a newly arriving job finds the full system, it joins
the orbit. The original primary jobs arrive to the system according
to a renewal process with rate $\lambda$. We denote the arrival
times of the original primary jobs by $t_n$ and we denote the
interarrival times by $\tau_n=t_{n+1}-t_n,\,n\ge 1,$ with generic
element $\tau$. Without loss of generality we assume that $t_1=0$.
The jobs have general i.i.d. service times $\{S_n^{(1)},\,n\ge 1\}$
with service rate $\mu$ and generic element $S^{(1)}$. Retrial
times $\{S_n^{(2)},\,n\ge 1\} $ are i.i.d. exponential with
(generic) service time $S^{(2)}$ with rate $\mu_0$ and independent
of the orbit size, provided it is positive. Such a model
is referred to as a retrial model with {\it constant retrial rate}. It
then follows that the orbit can be interpreted as a single-server
$\cdot/M/1$-type queue with service rate $\mu_0$ and with the input
which is formed by the flow of jobs rejected from the primary queue.
We use notation $\cdot/M/1$ because the merged stream arriving in
the orbit is not in general $GI$-type since it is a complex
combination of a lost part of original primary customers and
secondary customers returning to the orbit after unsuccessful
attempts to enter the primary queue.

In this work we establish {\it minimal sufficient stability
conditions} for the presented retrial system. Our model is quite
general. To the best of our knowledge,
the retrial queueing system with constant retrial rate and general renewal
arrival process is considered for the first time. In \cite{F86} Fayolle
has introduced a retrial system with
constant retrial rate. Fayolle has derived stability conditions for
the case of M/G/1/1 primary queue. In \cite{A96} Artalejo has
obtained stability conditions for the Markovian case M/M/2/2. In
\cite{RG98} Ramalhoto and G\'omez-Corral have obtained stability
conditions for the M/M/1/2 case. For the general Markovian case
M/M/c/K the authors of \cite{RG98} have obtained decomposition
results assuming ergodicity. The ergodicity conditions for the
multiserver Markovian case M/M/c/c with recovery probability have
been derived by Artalejo, G\'omez-Corral and Neuts in \cite{AGN01}.
In \cite{KKR06} Krishna Kumar and Raja have derived stability
conditions for the M/M/c/c constant retrial rate model with feedback and
balking. Stability conditions for the basic M/M/c/c constant retrial
model can be recovered from the results in \cite{AGN01} and
\cite{KKR06}. The ergodicity conditions for the Markovian case
M/M/1/K have been obtained in \cite{AY08}. In the works
\cite{AGN01}, \cite{AY08} and \cite{KKR06}  the authors have established
stability conditions by using the matrix-analytic technique for QBD processes \cite{N81}.
We shall demonstrate that stability conditions for M/G/1/1, M/M/c/c and
M/M/1/K systems are particular cases of our general conditions. Furthermore,
the following important cases with Poisson input have not
previously been covered: M/G/c/c (Erlang model), M/M/c/K (Markovian
model with a general number of servers and waiting spaces), M/D/1/K system.
Stability conditions of these important cases appear to be particular cases
of our general conditions and will be considered in detail below. We emphasize
that, to the best of our knowledge, such retrial system with general
renewal input of primary customers is considered for the first time.

There is a number of applications of retrial systems with constant retrial rates in
telecommunication. Using a retrial queue with constant retrial rate Fayolle \cite{F86}
has modelled a telephone exchange system.
In the series of papers \cite{CSA92}-\cite{CRP93} the authors have proposed to use a retrial
queueing system with constant retrial rate to model Multiple Access protocols. In particular,
in \cite{CSA92} the authors have modelled an unslotted Carrier Sense Multiple Access
with Collision Detection (CSMA/CD) protocol and in \cite{CPP93} and \cite{CRP93}
the authors have modelled some versions of the ALOHA protocol. In \cite{AY08} and \cite{AY10}
the authors have suggested to use retrial queues and retrial networks with constant retrial
rates to model TCP traffic originated from short HTTP connections.

The stability analysis used in this paper is based on renewal theory
and a characterization of the limiting behavior of the forward
renewal time in the process generated by regenerations of a basic
process. This approach, presented in  a general form  in
\cite{Morozov04, MorozovDelgado},  turns out to be effective in the
stability analysis of many queueing systems including general
multiserver retrial queue \cite{Morozov07}, and also multiserver
system with non-identical servers \cite{Morozov97}. The presented
method also works successfully outside of Markovian models, and
it is demonstrated also in this paper where non-Markov processes are
considered.  In particular, it allows us to
 reduce the dimension of the processes that simplifies analysis
considerably and does not require involved stability techniques
developed in the theory of multi-dimensional Markov processes
\cite{Borovkov,MEYN}.  An important contribution of this work is an
extension  of the stability analysis  to arbitrary initial state of
the system.

The paper is organized as follows. In the next Section~2 we present
the main result of the paper, minimal stability conditions for
GI/G/c/K-type retrial queue with constant retrial rate. We also
provide the proof of the main result. Then, in Section~3 we specify
the general condition for a number of important particular cases of
the primary queue. In particular, we consider the general Markovian
queue, the Erlang queue and the queue with deterministic service. We
would like to note that stability conditions for these important
particular retrial queues have not been available before. We
conclude our paper with Section~4.

\section{Stability analysis}
To describe the  behavior of the system, we  consider
(right-continuous) process $M(t):=N(t)+\nu(t),\,t\ge 0$, where
$N(t)$ is the number of retrial customers being in orbit and
$\nu(t)$ is the number of the customers {\it waiting in the buffer},
at instant $t$. Note that $\nu(t)\in [0,\,K-c]$ for any $t$. Also we
introduce the (right-continuous) process $W(t),\,t\ge0,$ expressing,
at each instant $t$, the {\it remaining workload in all servers}.
More exactly, if  $S_i(t)$ is the remaining service time at server
$i$ at instant $t$ ($=0$ if the server is empty), then
$W(t)=\sum_{i=1}^c S_i(t)$.
 Introduce the basic (two-dimensional) process
$X=\{X(t):=(M(t),\,W(t)),t\ge 0\}$. The  choice of the basic process
is motivated by the stability analysis  under arbitrary initial
state of the system, while there are other candidates for basic
process, to analyze the system under zero initial state. In the
latter case,  one can use, for instance,  the total number of
customers in the system, or the total remaining workload in the
system (including orbit), etc. Also, let $X(t_k^-)=X_k ,\,k\ge1$.
Denote $T_0=0$, then the instants
\begin{eqnarray}
 \quad T_{n+1}=\inf_k(t_k > T_n: X_k=(0,0)), \quad n\ge 0,
\label{1}
\end{eqnarray}
  are the  regeneration points of
the basic process $X$. Let  $T$ be a generic regeneration period and
$T(t)=\inf_k\{T_k-t: T_k-t>0\}$ be the  forward
renewal/regeneration time at instant $t\ge 0.$ If
\begin{eqnarray}
T_1<\infty\;\;\mbox{with probability 1 (w.p.1)  and }\;\;\; \E
T<\infty, \label{4}
\end{eqnarray}
then  we call any regenerative process having regeneration instants
defined by (\ref{1}) (and also the original system) {\it positive
recurrent}.  (This term, repeatedly used in previous works
\cite{Morozov97,Morozov04,Morozov07,Sig3, SW}, has evident analogy
with positive recurrent Harris Markov chains whose embedded renewal
process of regenerations has finite mean cycle length.)

It follows from the theory of regenerative processes that positive
recurrence is the most essential element in  stability analysis of
the process. Indeed, if the interarrival time $\tau$ is non-lattice
then $T$ is so, and, under positive recurrence, $X(t)$ converges to
a stationary limit. (It is obvious that  the system is unstable  if
$\E T=\infty$.) Our approach to stability is based on the following
result \cite {Feller}:  if $\E T=\infty$ then
\begin{eqnarray}
T(t) \Rightarrow \infty\,\,\,\,\mbox{ as}\,\,\,t\to \infty,\,\,
\label{3}
\end{eqnarray}
regardless of  initial (finite) delay $T_1=T(0)$, where
$\Rightarrow$ stands for convergence in probability. Thus, if
convergence (\ref{3}) does not hold then (\ref{4}) is satisfied and
positive recurrence for the {\it zero- delayed} process holds, in
which case $T_1=T$ and $X(0)=X_1=(0,0)$ (that is $t_1=0$ is a
regeneration instant). However, violation of (\ref{3}) does not
implies in general the finiteness of $T_1$ w.p.1 under arbitrary
initial state $X(0)$, and an extra (sometimes hard)  work is
required to prove that $T_1<\infty$. Denote  $\beta_0=0$, then
\begin{eqnarray}
\beta_{n+1}&=&\inf_k (k>\beta_n: X_k=(0,0)) ,\,n\ge 0, \label{112}
\end{eqnarray}
are  regeneration instants  of the embedded discrete-time  process $
X_n,\,n\ge1,$ with generic regeneration cycle length $\beta$, and
its remaining renewal time $\beta(n):=\inf_k\{\beta_k-n:
\beta_k-n>0\}\Rightarrow \infty $ (as $n\to \infty$) provided
$\E\beta=\infty$.

Now we outline how to apply the above mentioned approach to our
model. Because the buffer of the primary queue is finite, the source
of instability of the system can only be the unlimited increase of
the orbit size. Thus, we first show that under the predetermined
stability condition (see Theorem~2.1) the orbit size $N(t)\not
\Rightarrow\infty$.
Then
the second step is to show that also $T(t)\not \Rightarrow \infty$.
Finally, we apply  characterization (\ref{3}).
To establish $T_1<\infty$ for non-zero initial state, we will use
new development of the approach presented in \cite{MorozovDelgado}.

Let us denote original system by $\Sigma$ and construct an auxiliary
(new) system $\hat \Sigma$ as follows. The system $\hat \Sigma$ has
the same set of servers and the same buffer as system $\Sigma$, the
same renewal input of primary customers with rate $\lambda$ (we call
them $\lambda$-customers) and in addition {\it an independent
Poisson input of primary customers} with rate $\mu_0$ (we call them
$\mu_0$-customers). Arriving primary customer (of any type) who
finds servers and the buffer full in system $\hat \Sigma$  joins the
orbit. The orbit is an infinite buffer system of $\cdot/M/1$ type as
in the original system. The secondary customers leaving orbit (in
$\hat \Sigma$) {\it leave the system forever and do not affect its
future state}. Note that the system $\hat \Sigma $ (as $\Sigma$)
regenerates at the instants when the $\lambda$-customers find both
buffer and orbit empty. Note that here we also use the memoryless
property of the input of $\mu_0$-customers. Moreover, for any
variable $\zeta$ in system $\Sigma$ we denote corresponding variable
in system $\hat \Sigma$ as $\hat\zeta $. In particular,   $\hat W_k$
is  the remaining workload in all servers and $\hat\nu_k$ is  the
number of waiting customer in the buffer, respectively, in the
system $\hat \Sigma$ at  instant $t_k^-,\,k\ge1$.

Denote the primary queue of system $\hat \Sigma$ as $\hat \Theta$,
and note that it can be considered as an {\it isolated system}
because secondary customers (in $\hat \Sigma)$ leave the system and
{\it do not go to the primary queue}. Then the subsystem $\hat
\Theta$ regenerates (in continuous time) at the instants
 \begin{eqnarray}
\hat \Psi_{n+1}&=&\min(t_k>\hat \Psi_n: \hat W_k=\hat\nu_k =0),\,\,
\,n\ge 0,\label{55}
\end{eqnarray}
where, by definition, $\hat \Psi_0=0$. (So the regenerations of the
whole system $\hat \Sigma$ are a subsequence of the regenerations of
 $\hat \Theta$.)
Because the subsystem $\hat \Theta$ has a regenerative input and
finite buffer, then  such a system is positive recurrent under
the condition
\begin{eqnarray}
\P(\tau>S^{(1)})>0,\label{6}
\end{eqnarray}
see \cite{Morozov04, MorozovDelgado}.  In particular,
mean  generic regeneration period $\E \hat \Psi<\infty$.

\begin{rem} Exact  condition  in  \cite{Morozov04} (condition
(3.15) there) and in  \cite{MorozovDelgado} (condition (31) there)
 require  that the discrete-time regeneration period $\hat A$ (counting  all
 arrivals during regeneration period $\hat \Psi$) equals 1 with a positive probability.
It occurs if the interarrival time $\tau$  following a
$\lambda$-customer (with service time $S^{(1)}$) starting new
regeneration period is larger than $S^{(1)}$ and less than the next
Poisson arrival (with interarrival time $\tilde \tau$). In other
words, the following inclusion holds: $ \{\hat
A=1\}\supseteq\{\tilde \tau>\hat \tau>S^{(1)}\}.$ Obviously,  under
condition (\ref{6}), $\P(\tilde \tau> \tau>S^{(1)} )>0$, and  the
required assumption is fulfilled.
\end{rem}

\medskip

Denote by $\hat R(t)$ the total number of rejected customers in
system $\hat \Theta$ in the interval $[0,\,t]$ (this is also the total
number of customers which went to the orbit in the whole system
$\hat \Sigma$). Denote by $\hat A(t)$ the total number of arrivals
(primary $\lambda$-customers and $\mu_0$-customers) in the interval
$[0,\,t]$. Denote also by $\hat R$
the number of
rejected customers
during regeneration cycle  of system $\hat \Theta$.

Of course, the process $\{\hat R(t),\,t\ge 0\}$ is positive
recurrent cumulative process with  embedded regenerations $\{\hat
\Psi_n\}$ and, in particular,
there exists
(w.p.1) the limit
\begin{eqnarray}
\lim_{t\to \infty}\frac{\hat R(t)}{\hat A(t)}= \frac{\E \hat R}{\E \hat
A}. \label{5}
\end{eqnarray}
(To explain, we note that  $\hat R (t)/t\to \E \hat R/\E\hat
\Psi,\,\hat A(t)/t\to \E \hat A/\E\hat \Psi$.) In the system $\hat
\Sigma$, define indicator $I_n$ as
\begin{equation}
\label{InDef} I_n = \left\{ \begin{array}{ll}
1, & \mbox{if customer $n$ is rejected,}\\
0, & \mbox{otherwise},
\end{array}\right.
\end{equation}
so the sequence $\{I_n,\,n\ge 1\}$
has regeneration period $\hat A$. Because $\P(\hat A=1)>0$, then the
weak limit $I_n\Rightarrow I$ exists, or $\P(I_n=1)\to \E
I:=\P_{loss}$
where $\P_{loss}$ is stationary loss
probability. Moreover,
by the standard result of regenerative theory, stationary loss
probability coincides with the long-run-average loss probability
(\ref{5}):
\begin{eqnarray}
\P_{loss}=\lim_{n\to \infty}\frac{\sum_k^n I_k}{n}=\frac{\E \hat
R}{\E \hat A}.\label{loss}
\end{eqnarray}

\noindent Now we are ready to formulate the main stability result.

\begin{thm}\label{thm:stability} Assume that
condition $(\ref{6})$ and the following condition
\begin{eqnarray}
(\lambda+\mu_0)\P_{loss}<\mu_0, \label{7}
\end{eqnarray}
hold.  Then, under  arbitrary (fixed) initial state $X (0)=
X_1=(m_0,\,w_0)$, the original system is positive recurrent, that is
$$
\E T<\infty,\,\E \beta<\infty\;\; \mbox{and}\;\;
T_1<\infty,\;\mbox\,\beta_1<\infty \;\;\mbox{w. p. 1}.
$$
Moreover, the stationary  distribution $\lim_{n\to
\infty}\P(X_n\in\cdot)
$ exists. If, in addition, interarrival time $\tau$ is non-lattice,
then  stationary distribution $\lim_{t\to \infty}\P(X(t)\in\cdot)$
also exists.
\end{thm}

{\it Proof.} We will use a monotonicity property of the loss system
with respect to change of service times, see \cite{Sonderman1,
Sonderman2, Whitt}. Namely, we use coupling to sample identical
corresponding interarrival times in both systems
$\Sigma,\,\hat\Sigma$ until first empty period of  the orbit  in the
original system $\Sigma$. Then we assume that during this (and any
following) empty period of the orbit we continue to sample input
Poisson process (with parameter $\mu_0$) in the original system but
with arriving customers having {\it zero service times} (as long as
the empty period lasts). At the end of such period, we interrupt
current interarrival times (in  Poisson processes in both systems
$\Sigma$ and $\hat \Sigma$) and resample new (identical)
interarrival times for both systems. By memoryless property, this
resampling keeps distribution of the input process. At the same time
this procedure allows us to keep  equivalence between input
intervals in both systems. (Note that $\lambda$-inputs stay
identical and unchanged in both systems.)   Moreover, this shows
that service times of arriving customers in the systems are
stochastically ordered as $S_n^{(1)}\le_{st}\hat  S_n^{(1)}$
while for non-zero (actual) customer $n$, $S_n^{(1)}=_{st}\hat
S_n^{(1)},\,n\ge 1$.

Now we can use the result of \cite{Sonderman1} which claims (in
adaptation to our model) that if two (the same) finite capacity
systems have the same input and the ordered service times, as above,
then the number of rejected customers in primary queues of systems
$\Sigma$ and $\hat\Sigma$, respectively, in the interval $[0,\,t]$
are ordered as
\begin{eqnarray}
R(t)\le_{st} \hat R(t),\,\,t\ge 0. \label{8}
\end{eqnarray}
We emphasize that $R(t)$  counts  all rejections which  happen
in  system $\Sigma$ including repetitive rejections of the orbit
customers after unsuccessful attempts to enter the primary queue.
Because the input in system $\hat\Sigma$ is a superposition of two independent renewal
processes, then w.p. 1 as $t\to\infty$
 \begin{eqnarray}
\frac{A(t)}{t}\to \lambda+\mu_0.
\end{eqnarray}
Since $\hat R(t)=\sum_k^{A(t)}I_k$, it then follows from (\ref{5}) and
(\ref{loss})
 that w.p.1
 \begin{eqnarray}
\lim_{t\to \infty}\frac {\hat R(t)}{t}=
(\lambda+\mu_0)\P_{loss}.\label{20a}
\end{eqnarray}

(It follows from $\hat R(t)\le A(t)$ and  $\E A(t)/t\to
\lambda+\mu_0$,
that the family $\{\hat R(t)/t,\,t\ge 0\}$
is uniformly integrable, and thus
the limit
$\E \hat R(t)/t\to   (\lambda+\mu_0)\P_{loss}$
also  exists.)

Denote by $V_o (t),\,\hat V_o(t)$ the total workloads arrived {\it
to the orbit} in the systems $\Sigma,\,\hat\Sigma$, respectively,
during the interval $[0,\,t]$, including the same (arbitrary) initial state
$V_o(0)=\hat V_o(0):=V_o$.
Note that
 \begin{eqnarray}
\hat V_o(t)=V_o(0)+\sum_{k=1}^{\hat R(t)}S_k^{(2)},
 \,\,\,t\ge 0\;\;\;(\sum_\emptyset=0),
\end{eqnarray}
and thus,
\begin{eqnarray}
\frac{\hat V_o(t)}{\hat R(t)}\to  \frac{1}{\mu_0}\;\;   \mbox{w.p.1
as}\;t\to \infty.\label{22-1}
\end{eqnarray} Denote, in the original system $\Sigma$, by $\mu_o(t)$ the
total empty time of the orbit  in interval $[0,\,t]$ and by
$W_o(t)$  the (right-continuous) remaining  workload in orbit  at
instant $t\ge 0$. Now we have the following balance equation
\begin{eqnarray}
V_o(t)=W_o(t)+t-\mu_o(t),\;\;t\ge 0.\label{balance}
\end{eqnarray}
Thus, we have
\begin{eqnarray}
\mu_o(t) &\ge&
t-V_o(t)=t-\sum_{k=1}^{R(t)}S_k^{(2)}-V_o(0)\nonumber\\
&\ge_{st}& t-\sum_{k=1}^{\hat R(t)}S_k^{(2)}-\hat V_o(0)=t-\hat
V_o(t),\,\,\,t\ge 0.\label{22a}
\end{eqnarray}
 By  (\ref{7}), (\ref{20a}),
 (\ref{22-1})
 this implies
\begin{eqnarray}
\liminf_{t\to\infty}\frac{\mu_o(t)}{t}\ge
1-\frac{(\lambda+\mu_0)\P_{loss}}{\mu_0}:=\delta_0>0.\label{20}
\end{eqnarray}
 This also shows that under arbitrary initial state $V_o(0)$,
\begin{eqnarray}
\mu_o(t)=\int_0^tI(N(u)=0)du\to \infty \;\;\mbox{w.p.1 as}\;\;t\to
\infty,\label{22}
\end{eqnarray}
where $I$ denotes indicator function. By Fatou's lemma,
\begin{eqnarray}
\liminf_{t\to \infty}\frac{1}{t} \E \mu_o(t)>0.
\end{eqnarray}
Thus,  $\P(N(t)=0)\not\to 0$  as $t\to\infty$ and there exist
$\delta^*>0$ and non-random sequence of instants $z_n\to \infty$ such that
\begin{eqnarray}
\inf_{n\ge 1}\P (N(z_n)=0)\ge \delta^*.\label{delta*}
\end{eqnarray}
Starting from this point we work with the original system
$\Sigma$ only. Let  $\tau(t)=\min_k(t_k-t: t_k-t\ge 0)$ be the
remaining interarrival time at instant $t$ in system $\Sigma$. Note
that, for each $z_n$,
\begin{eqnarray}
\P(N(z_n)=0)=\P\Bigl(N(z_n+\tau(z_n))=0\Bigr)\ge \delta^*.
\end{eqnarray}
Denote $\kappa(n)=\min(k: t_k\ge z_n)$, then
$z_n+\tau(z_n)=t_{\kappa(n)}$ is the first arrival instant after
$z_n,\,n\ge 1$. Let $L(t)$ be the total  remaining
workload at the servers and buffer at instant $t$. Obviously,
\begin{eqnarray}
L(t)\le _{st}
W(t)+\sum_{i=1}^{K-c}S_i^{(1)},\,\,t\ge0.
\end{eqnarray}
 Note that  the remaining
service time processes, $\{S_i(t) ,\,t\ge 0\},\,i=1,\ldots,c,$ are
tight. (To have the tightness, the only requirement, besides the
finiteness of  $\E S^{(1)}$, is that service times $\{S_n^{(1)}\}$
are independent of the input process \cite{Morozov97}.) Thus, the
process $W(t),\,t\ge 0,$ is tight, and hence, the process
$L(t),\,t\ge 0,$ is also tight. Then  we can find a constant $D$
such that $\inf_n \P(L(z_n)\le D)\ge 1- \delta^*/2.$  Now we fix for
a moment some $z_n$ satisfying (\ref{delta*}) and let
$L(t_n^-)=L_n,\,n\ge1$. Then, we have
\begin{eqnarray}
\P(N_{\kappa(n)}=0,\,L_{\kappa(n)}\le D)\ge \P(N(z_n)=0,\,L(z_n)\le
D)\ge \frac{\delta^*}{2},\label{27}
\end{eqnarray}
where the first inequality is valid since the workload decreases in
interval $[z_n,\,t_{\kappa(n)})$ and the second inequality follows from
$$
\P(N(z_n)=0,\,L(z_n) \le D) = \P(L(z_n) \le D) - \P(N(z_n)>0,\,L(z_n) \le D)
$$
$$
\ge \P(L(z_n) \le D) - \P(N(z_n)>0).
$$
Next we introduce the event
$${\cal
E}_n:=\{N_{\kappa(n)}=0,\,L_{\kappa(n)}\le D,\,\tau(z_n)\le C\}
$$
where, by the tightness of the process $\{\tau(t),t\ge 0\}$, the
constant $C$ is taken in such a way that $\P({\cal E}_n) \ge
\delta^*/4$. Thus,  on the event  ${\cal E}_n$, the  customer
$\kappa(n)$
indeed arrives in interval $[z_n,\,z_n+C]$ and
finds  the workload $L_{\kappa(n)}\le D$. It follows from (\ref{6})
and $\E \tau<\infty$ that one can find constants
$a<\infty,\,\gamma>0,\,\epsilon>0$ such that
 \begin{eqnarray}
\P(a\ge \tau>\gamma+S^{(1)})=\epsilon.\label{30}
\end{eqnarray}
Note that, for each $i$, on  the event $\omega_i:=\{ a\ge
\tau_i>\gamma+S_i^{(1)}\}$
 the workload accumulated in the system at instant $t_i$, decreases
during interarrival time $[t_i,\,t_{i+1})$ not less than by
$\gamma$, {\it provided  that at least one server is not empty}
during this interval.
If only orbit is not empty,  then
 an orbit
customer may attempt  to enter server/buffer before next arrival
instant $t_{i+1}$. In such a case, regardless of the retrial attempt
was successful or not, on the event $\omega_i$, the accumulated
workload decreases not less than by $\gamma/2$ (during
$[t_i,\,t_{i+1})$) with the probability $\ge
\epsilon(1-e^{-\mu_0\frac{\gamma}{2}})$. Denote $R:=\lceil 2
D/\gamma \rceil$. Then on the event
$$
{\cal E}_n\bigcap\bigcap_{i=\kappa(n)}^{\kappa(n)+R} \omega_i
$$
$R$ primary customers
arrive in the interval $[z_n,\,z_n+C+aR]$, and the accumulated workload reaches
zero (regeneration) with probability $\ge \frac{1}{4} \delta^* \epsilon^R
(1-e^{-\mu_0\frac{\gamma}{2}})^R>0.$ Because this lower bound is
uniform in $n,\,z_n$, then  $\E \beta<\infty$. It follows from
Wald's identity and representation $T=_{st}\sum_{k=1}^\beta\tau_k$
that mean cycle period  for continuous time processes is finite, $\E
T=\E \beta \E \tau <\infty$.

Using a modification of the approach from \cite{MorozovDelgado}, we
now  extend  analysis  to arbitrary initial state of the basic
process $X$. Let $S_{k(i)}^{(1)} $ be the $k$-th service time
realized at server $i$. (For each $i$, these service times are
i.i.d. and distributed as $S^{(1)}$.) Introduce
$$
\tilde
S_i(t)=\min_{k(i)}(S_{1(i)}^{(1)}+\cdots+S_{k(i)}^{(1)}-t:\,S_{1(i)}^{(1)}+\cdots+S_{k(i)}^{(1)}-t>0)
$$
the remaining  renewal time at instant $t$ in the renewal process
generated by service times of  server $i$, and let $\tilde
S_i(0)=S_i(0),\,i=1,\ldots,c$. For any (integer) $r\ge 0$ and $x\ge
0$,  denote the set $\verb"B"(r,\,x)=[0,\,r]\times[0,\,x]$ and
consider the process $Y=:\{Y(t)=(N(t),\,W(t)),\,t\ge 0$\}. Then, for
any $t$ and $x$:
\begin{eqnarray}
I( Y(t)\in \verb"B"(0,\,x))
&\ge &I(N(t)=0)-I(W(t)>x)\nonumber\\
&\ge& I(N(t)=0)-\sum_{i=1}^c I\Bigl(S_i(t)>\frac{x}{c}\Bigr).
\nonumber\\\label{33}
\end{eqnarray}
Although,  for each $t$, $S_i(t)$ and $\tilde S_i(t)$ in general are
not comparable,  but  by construction, the following inequality
holds for any $x$:
\begin{eqnarray}
\int_0^tI\Bigl(S_i(u)>x
 \Bigr)du\le \int_0^tI\Bigl(\tilde S_i(u)>x
\Bigr)du,\,\,i=1,\ldots,c. \label{34}
\end{eqnarray}
Note that $I\Bigl(\tilde S_i(t)
>x
\Bigr),\,t\ge0, $ is a regenerative process with  (generic) cycle
length $S^{(1)}$. Then w.p.1 as $t\to \infty$,
\begin{eqnarray}
\frac{1}{t} \int_0^tI\Bigl(\tilde S_i(t)>x \Bigr)\to \frac{1}{\E
S^{(1)}}\int_x
^\infty\P(S^{(1)}>t)dt:=\P(S_e^{(1)}>x),\,\,\,i=1,\ldots,c,
\label{35}
\end{eqnarray}
where distribution of the stationary overshoot  $S_e^{(1)}$ is
proper. Hence,  we can take $x_1$  in such a way that
\begin{eqnarray}
\P\Bigl(S_e^{(1)}>\frac{x_1}{c}\Bigr)<\frac{\delta_0}{2c}.
\label{36}
\end{eqnarray}
If we take $x=x_1$ in (\ref{33})-(\ref{35}) then it follows from
(\ref{20}), (\ref{33})--(\ref{36}) that w.p.1
\begin{eqnarray}
\liminf_{t\to \infty}\frac{1}{t}\int_0^tI (Y(u)\in \verb"B"
(0,\,x_1))du\ge \liminf_{t\to
\infty}\frac{\mu_o(t)}{t}-\frac{\delta_0}{2}=\frac{\delta_0}{2}.\label{30}
\end{eqnarray}
In particular,  the total time that the process $Y$ spends in the
set $\verb"B"(0,\,x_1)$ within time interval $[0,\,t]$
\begin{eqnarray}
\int_0^tI(Y(u)\in \verb"B"(0,\,x_1)) du\to \infty,\,\,\,\,t\to
\infty.\label{31}
\end{eqnarray}
Because,  for each $t$ and $x\ge 0$,
\begin{eqnarray}
\{Y(t)\in \verb"B"(0,\,x)\}\subseteq \{X(t)\in \verb"B" (K-c,
\,x)\},
\end{eqnarray}
then, by (\ref{31}), the total time  the process $ X$  spends in the
set $\verb"B"(K-c,\,x_1)$ during time interval $[0,\,t]$
$$
\mu_B(t):=\int_0^tI(X(u)\in \verb"B" (K-c, \,x_1)) du\to
\infty\;\,\,\mbox{w.p.1 as}\;\;t\to \infty.
$$
Denote by $\Lambda(t)$ the number of $\lambda$-customers arriving
in the interval $[0,\,t]$, and let $G_B(t)$ be the number  of these
$\lambda$-customers
 who meets the process  $ X$
 in the set $\verb"B" (K-c, \,x_1)$, that is
\begin{eqnarray}
G_B(t)=\sum_{k=1}^{\Lambda(t)} I( X_k\in\verb"B" (K-c, \,x_1)).
\label{end}
\end{eqnarray}
It  then follows  that  w.p.1
\begin{eqnarray}
G_B(t)(\max_{1\le n\le \Lambda(t)}\tau_n)\ge \mu_B(t)\to
\infty,\,\,t\to \infty. \label{eq:F5c}
\end{eqnarray}
 Since  $\E\tau<\infty$,  then
$\max_{1\le n\le \Lambda(t)}\tau_n=o(t),\,\,\, t\to \infty$ \cite
{Smith}. Thus, $G_B(t)\to \infty$ w.p. 1 as $t\to\infty$, and the
total number of $\lambda$-customers, $G_B(\infty)$, meeting the
process $ X $ in the set $\verb"B" (K-c, \,x_1)$ is infinite w.p. 1.
(This holds for arbitrary initial state $X(0)$, see
(\ref{20}),(\ref{22}),(\ref{30}).) Recall initial state
$X(0)=(m_0,\,w_0)$, and    choose and fix arbitrary $r\ge
\max(m_0,\,K-c)$ and $z \ge \max (w_0,\,x_1)$. Then, as in
\cite{MorozovDelgado}, one can find constants $D_0<\infty,
\,\varepsilon_0>0$ such that the remaining renewal time $\beta(k)$
satisfies
\begin{eqnarray}
\inf _{k\ge 1}\P_{(m_0,\,w_0)}(\beta(k)\le D_0\,|\, X_k\in \verb"B"
(r, \,z),\,\beta_1>k)\ge \varepsilon_0,\label{35}
\end{eqnarray}
where  the bounds $\varepsilon_0$ and $D_0$ are uniform:  they
depend on $r,\,z$ but  does not depend on neither customer number
$k$ nor the specific states $X_k
\in \verb"B" (r, \,z)$. Indeed, due to the specific form of the
basic process $X$, conditioned on the event $\{ X_k\in
\verb"B" (r, \,z),\,\beta_1>k\}$, the total
workload in the system at instant $t_k^-$ is (uniformly in $k$)
stochastically upper bounded  by the quantity
\begin{eqnarray}
z+\sum_{i=1}^{K-c}S_i^{(1)}+\sum _{i=1}^rS_i^{(2)},
\end{eqnarray}
where, evidently,   both the service times $\{S_i^{(1)}\}$ of
waiting customers in the buffer and (exponential) retrial times
$\{S_i^{(2)}\}$ of all orbit customers  are i.i.d
 and independent of the remaining workload $W_k\,(\le z)$ in the
servers.   Then one can unload the system during $D_0$
$\lambda$-arrivals with a  probability $\ge \varepsilon_0$ (like in
the proof after formula (\ref{30})), and thus condition (\ref{35})
is indeed satisfied.  Then one can show that  the mean number
of $\lambda$-customers meeting the process $X$ in the set $\verb"B"
(r, \,z)$ during first regeneration period is finite,
\begin{eqnarray}
\E _{(m_0,\,w_0)}\Bigl( \sum_{k=1}^\beta I(X_k\in \verb"B" (r,
\,z))\Bigr) \le \frac{D_0}{\varepsilon_0}<\infty.\label{37}
\end{eqnarray}
 (More details can be found in
\cite{Morozov07, MorozovDelgado}.)  Because $\verb"B" (r,
\,z)\supseteq \verb"B" (K-c, \,x_1)$, then (\ref{37}) implies that
the number of $\lambda$-customers meeting the process $X$ in the set
$\verb"B" (K-c, \,x_1)$ during first regeneration period is finite,
$G_B^{(1)}:=\sum_{k=1}^\beta I(X_k\in \verb"B" (K-c,
\,x_1))<\infty$.
Then it follows from  above that
$\P_{(m_0,\,w_0)}(G_B^{(1)}<G_B(\infty))=1$, and hence the number of
regeneration cycles is not less than two. Thus, first regeneration
period $\beta_1<\infty$ and hence
$T_1=\tau_1+\cdots+\tau_{\beta_1}<\infty$.
(Indeed, the number of regeneration cycles is infinite w.p.1.)
Finally, we note that condition (\ref{6}) implies
aperiodicity  of $\beta$.
 \vspace{3 mm} \rule {1.5ex}{1.5ex}
\bigskip

\begin{rem}
In general, one can define another type of regeneration, for
instance,  $k$-{\it regeneration}, when an arriving $\lambda$-type
customer see all servers and buffer empty
and a fixed $k$ customers in the orbit, see \cite{Morozov07}.
Moreover, a relaxation (or elimination) of assumption (\ref{6})
(which is necessary to have classical regenerations
 of type (\ref{1}), (\ref{112}))
  may lead to the so-called  one-dependent (or
 weak)  regeneration with a dependence between adjacent
 regeneration cycles, see for instance, \cite {Morozov04, Sig3,SW}.
However, we do  not consider such regenerations in this work.
\end{rem}

\begin{rem}
In a quite general case of unbounded $\tau$ (and in particular for
the Poisson $\lambda$-arrival process), the constant $a$ in
(\ref{33}) can be taken in such a way that regeneration occurs in
the interval $[z_n,\,z_n+C+a]$.
\end{rem}

\section{Application to particular queueing models}

In this section we apply general stability conditions of Theorem~\ref{thm:stability}
to important particular types of the primary queue.

\subsection{Markovian case M/M/c/K}

In the Markovian case with $c$ servers, service rate $\mu$ and $K$ places in the primary
queueing system, the loss probability in the auxiliary $\Sigma_2$ system is given by
$$
P_{loss} = \frac{((\lambda+\mu_0)/\mu)^c}{c!} \left(\frac{(\lambda+\mu_0) /\mu}{c}\right)^{K-c} P_0,
$$
where
$$
P_0=
\left[\sum_{n=0}^c \frac{((\lambda+\mu_0) /\mu)^n}{n!} +
\frac{((\lambda+\mu_0) /\mu)^c}{c!} \sum_{n=1}^{K-c}\left(\frac{(\lambda+\mu_0) /\mu}{c}\right)^n \right]^{-1}.
$$
Hence, the stability condition (\ref{7}) takes the form
\begin{equation}
\label{GenerMark}
\frac{((\lambda+\mu_0) /\mu)^K}{c!} \left(\frac{(\lambda+\mu_0) /\mu}{c}\right)^{K-c} P_0 < \frac{\mu_0}{\lambda+\mu_0}.
\end{equation}
The other conditions of Theorem~\ref{thm:stability} are naturally satisfied by the Poissonian arrival process.
In the case of one server (M/M/1/K case), the above condition reduces to
$$
((\lambda+\mu_0)/\mu)^K \left[ \sum_{n=0}^{K} ((\lambda+\mu_0)/\mu)^n \right]^{-1} < \frac{\mu_0}{\lambda+\mu_0},
$$
which is equivalent to the ergodicity condition provided in \cite{AY08}.
Then, in the M/M/c/c case, the condition (\ref{GenerMark}) reduces to
\begin{equation}
\label{ccase}
\frac{((\lambda+\mu_0) /\mu)^c}{c!}
\left[\sum_{n=0}^c \frac{((\lambda+\mu_0) /\mu)^n}{n!}\right]^{-1} < \frac{\mu_0}{\lambda+\mu_0},
\end{equation}
which is equivalent to the ergodicity condition provided
in \cite{AGN01} with the recovery probability $p=1$.

It is important to note that the conditions given above are in
fact necessary and sufficient stability conditions for the
Markovian case. This is why we call conditions presented in
Theorem~\ref{thm:stability} minimal sufficient stability conditions.

\subsection{Erlang model}

When the primary queue is described by the Erlang model (M/G/c/c case), the general stability
condition (\ref{7}) takes the form
$$
\frac{\Bigl((\lambda+\mu_0) \E S^{(1)}\Bigr)^c}{c!}
\left[\sum_{n=0}^c \frac{\Bigr((\lambda+\mu_0) \E S^{(1)}\Bigr)^n}{n!}\right]^{-1} < \frac{\mu_0}{\lambda+\mu_0},
$$
which is the same form as in (\ref{ccase}). Thus, the stability
conditions demonstrate insensitivity property when the primary queue
is the Erlang queue. We would like to emphasize that it seems to be
very difficult (if possible at all) to obtain the above stability
condition using stability techniques for Markov chains and embedded
Markov chains. Anyway, the presented approach ensures short
and simple proof of stability.

\subsection{Deterministic service}

We can obtain explicit stability conditions in another important
case, in the case of deterministic service (M/D/1/K model for the
primary queue). In the case of deterministic service, the loss
probability is given by
$$
P_{loss} = 1-\frac{b_{K-1}}{1+\rho b_{K-1}},
$$
where
$$
b_{K-1}=\sum_{n=0}^{K-1} \frac{(-1)^n}{n!} (K-1-n)^n \rho^n e^{(K-1-n)\rho},
$$
with $\rho=\lambda S$ \cite{BG00}.  Thus, in the M/D/1/K case the
stability condition (\ref{7}) takes the form
$$
\frac{\lambda}{\lambda+\mu_0} < \frac{b_{K-1}}{1+\rho b_{K-1}},
$$
and we note that there exists an efficient recursive approach
to calculate the coefficient $b_{K-1}$ \cite{BG00}.

\section{Conclusion}

We have considered a retrial queueing system with general renewal
arrival process, general service time and constant retrial rate. To
the best of our knowledge a retrial queueing system with constant
retrial rate was considered for the first time under so general
setting. We have obtained minimal stability conditions which are
necessary and sufficient in the Markovian case. Stability
analysis also covers  arbitrary initial state of the system. The
conditions have clear probabilistic interpretation and can be easily
applied to a number of important particular cases. Examples of such
particular cases are Erlang model and deterministic service model.
We have observed that the stability condition for the Erlang model
with retrial is insensitive to the distribution shape.

\section*{Acknowledgements}

We would like to acknowledge the financial support provided by EGIDE
ECO-NET grant no.18933SL and RFBR grant no.10-07-00017.

\addcontentsline{toc}{section}{\textit{References}}
\bibliographystyle{unsrt}


\end{document}